\documentclass{appolb}
\usepackage{epsfig}
\usepackage{amsmath,amssymb}
\usepackage{slashed}
\def\runhead{}
\begin{document}
 \eqsec  % uncomment this line to get equations numbered by (sec.num)
\title{Neutrinos and Lepton-Flavour--Violation 
\thanks{Presented at FLAVIAnet Topical Workshop ``Low energy constraints on extensions of the Standard Model'',  Kazimierz, Poland, 23-27 July 2009.}%
% you can use '\\' to break lines
}
\author{Thorsten Feldmann
\address{Physik Department T31, Technische Universit\"at M\"unchen,\\ 
D-85747 Garching, Germany.}
%\and
%the Name(s) of other Author(s)
%\address{and their affiliation}
}
\maketitle
\begin{abstract}
I give a sketchy overview on aspects related to
the lepton flavour sector in the standard model and its possible 
extensions.
\end{abstract}
\PACS{14.60.Pq, 14.60.St, 13.35.-r.\hfill Preprint: TUM-HEP-734/09}
  
\section{Introduction}

The standard model (SM) of particle physics successfully describes
a variety of phenomena in terms of a few fundamental quantities. 
Besides the gauge coupling constants  associated to the
fundamental interactions, and the scalar
sector responsible for spontaneous electroweak symmetry breaking, 
the major part of adjustable parameters are related to the fermion sector
describing the masses and mixings of quark and lepton flavours. 

The experimental observation of neutrino oscillations
implies that neutrinos 
have (tiny) masses, and the SM -- which in its original version
is restricted to massless neutrinos -- has to be extended.
The most popular approaches are based on ``see-saw''
scenarios, where neutrino masses are suppressed by a large
scale which originates from integrating out heavy particles.
In particular, heavy Majorana neutrinos appear naturally when embedding
the SM in grand unified theories (GUTs). 
The violation of lepton number $L$ at the Majorana mass scale
allows to explain the baryon asymmetry in the universe 
by leptogenesis, where an $L$-asymmetry is built by out-of-equilibrium decays
of Majorana neutrinos, which is subsequently transformed into a baryon
asymmetry by electroweak sphaleron processes. The required CP-violation
enters through intereference effects of tree-level and loop diagrams
and depends on the phases in the Majorana sector.

While the above phenomena relate the lepton flavour sector to high-energy
scales and physics near the GUT scale, low-energy observables from
lepton-flavour violating (LFV) processes are very sensitive to
new physics (NP) at the TeV scale. For instance,  in the minimally extended SM 
(see below), the decay $\mu\to e\gamma$ has a tiny branching fraction,
${\cal B}[\mu \to e\gamma]_{\rm SM} \sim 10^{-54}$ 
compared to expectations from 
generic NP models which may be close to the present (foreseen) experimental reach,
 ${\cal B}[\mu \to e\gamma]_{\rm exp.} < 10^{-11(13)}$.
For more comprehensive discussions and more references to the original
literature, I refer to the recent reviews 
\cite{Strumia:2006db,Valle:2006vb,Mohapatra:2005wg,GonzalezGarcia:2007ib,Raidal,Schwetz:2008er}.

\section{Massive neutrinos and lepton-flavour mixing}

\subsection{Extending the SM lepton sector}

In the (original) SM, only charged leptons obtain a mass
from the Yukawa coupling to the Higgs field $H$,
\begin{align}
 (Y_E)^{ij} \ (\bar L^i \, H) \, E_R^j \ + \ \mbox{h.c.} 
\end{align}
Here $L$ and $E_R$ denote the three families of left-handed lepton doublets
and right-handed lepton singlets, respectively, and $Y_E$ is the corresponding
Yukawa coupling matrix.
The Lagrangian describes massless neutrinos with 
\emph{individual} lepton flavour ($L_e,L_\mu,L_\tau$) being conserved (i.e.\ no mixing).
Adding right-handed Dirac neutrinos $\nu_R$,
and enforcing $L$-conservation,
one obtains the analogous situation to the quark sector
(i.e.\ CKM-like mixing). In view of the observed qualitative differences
between the lepton and the quark sector, such a scenario
seems to be less appealing.

Considering, instead, the SM as an effective theory, one can obtain
neutrino masses in a minimal way
by including the dimen\-sion-5 operator
\begin{align} 
     \frac{(g_\nu)^{ij}}{\Lambda_{\slashed L}} \ 
     (\bar L^i \widetilde H)  (\widetilde H^\dagger L^j)^{c} + \mbox{h.c.} 
\label{dim5}
\end{align}
which violates lepton number, with the associated high-energy scale
denoted as $\Lambda_{\slashed L}$.
The mismatch between the diagonalization of the 
Yukawa matrix $Y_E$ and of the new flavour matrix 
$g_\nu=g_\nu^T$ yields the Pontecorvo-Maki-Nakagawa-Sakata (PMNS) mixing matrix.

Another option is to introduce right-handed Majorana neutrinos $\nu_R$ via
\begin{align}
      (Y_\nu)^{ij} \, (\bar L^i \, \widetilde H \, \nu_R^j) + 
     \frac12 \, M^{ij} \, (\nu_R^T)^i \, (\nu_R)^j +   \mbox{h.c.} 
\end{align}
In this case, lepton number is violated by the Majorana mass term $M$,
and one can reproduce the dim-5 term in (\ref{dim5}) by integrating
out $\nu_R$, 
\begin{align}
\frac{g_\nu}{\Lambda_{\slashed L}} = Y_\nu \, 
 (M)^{-1}\,  Y_\nu^T  \,,
\label{rel}
\end{align}
realizing the so-called type-I see-saw mechanism.
Alternative mechanisms introduce heavy scalar triplets (type-II see saw),
or heavy fermion triplets (type-III see saw).

\begin{table}[t] 
\caption{Parameter counting for the minimally extended SM. \label{tab:dim5}}
\begin{center} 
\begin{tabular}{lcrr}
\hline \hline 
Quantity & Symbol & Moduli & Phases \\
\hline &&& 
\\[-0.8em]
charged Yukawa matrix & $Y_E$ & 9 & 9 
 \\[0.2em]
dim-5 neutrino matrix & $g_\nu=g_\nu^T$  & 6 & 6  
\\[0.2em]
flavour symmetry group & $U(3)_L \times U(3)_{E_R}$
  &  $-6$ & $-12$ 
\\[0.2em]
 \hline 
\\[-0.8em]
physical parameters: & masses $m_\ell^i$, $m_\nu^i$  
 & $6$
\\[0.2em]
 & angles \& phases  
 & $3$
%\\[0.2em]
%& CP phases: & 
& $3$ 
\\[0.2em]
\hline \hline
\end{tabular}
\end{center}
\end{table}

In the minimally extended SM (\ref{dim5}), the parameter counting for the 
lepton-flavour sector follows from the symmetries broken
by the Yukawa sector \cite{Santamaria:1993ah}, see Table~\ref{tab:dim5}.
The low-energy neutrino-mixing parameters in the
PMNS matrix are thus given by
3 angles, 1 Dirac phase and 2 Majorana phases.
As has been pointed out (e.g.\ \cite{Broncano:2002rw}),
see-saw scenarios in general contain additional flavour parameters, which may not
be directly accessible at low energies.
For example, with 3 heavy Majorana neutrinos, the high-energy theory contains
3 additional angles and 3 additional phases, which can
be parametrized in terms of an orthogonal complex matrix 
\cite{Casas:2001sr},
such that in a basis where the charged-lepton Yukawa
matrix $Y_E$ and the Majorana mass matrix $M$ is diagonal, one has
\begin{align}
& Y_\nu^T = {\rm diag}[\sqrt M_\nu] \, R \,
 {\rm diag}[\sqrt m_\nu] \, U^\dagger_{\rm PMNS} \, / \, \langle H \rangle \,.
\end{align}
The matrix $R$ drops out in the dim-5 coefficient matrix $g_\nu$
in (\ref{rel}), but will contribute to operators of dim-6 or higher.
If $\Lambda_{\slashed L} \sim M_{\rm GUT}$, the coefficients of the latter
will be highly suppressed, while generic $L$-conserving NP effects,
associated with a scale
$\Lambda_{\rm LFV} \gtrsim 1$~TeV, would clearly dominate.

%%%%%%%%%%%%%%%%%%%%%%%%%%%%%%%%%%%%%%%%%%%%%%%%%%%%%%%%%%%%%%%%

\subsection{Experimental situation}

The experimental determination of neutrino-mixing parameters
(see e.g.\ \cite{Schwetz:2008er,Raidal} and references therein) 
reveals: 
 \begin{itemize}
  \item two distinct mass-squared differences, 
       $|\Delta m_{31}^2| \sim 2 \cdot 10^{-3}~{\rm eV}^2$ related to atmospheric
       neutrino oscillations, and $\Delta m_{21}^2 \sim 7 \cdot 10^{-5}~{\rm eV}^2$ to solar neutrino oscillations;
  \item a mixing angle, $\sin^2\theta_{23} \simeq 0.5$, close to being maximal;
%  \item 
  a large mixing angle $\sin^2\theta_{12} \sim 0.3$;
%  \item 
and a small mixing angle $\theta_{13}$.
 \end{itemize}
Current data still leave open whether neutrinos
have a normal/inverted hierarchical or a degenerate spectrum. 
The Majorana nature has to be verified/falsified,
for instance by searching for $\nu$-less double $\beta$-decay. 
Together with 
constraints from cosmology and from
the endpoint of $\beta$-decay energy spectra,
this also helps to set the absolute neutrino mass scale.
A precise measurement of  $\theta_{13}$ is foreseen in
the near future, whereas the determination of CP-phases  will 
be difficult (see e.g.\ the discussion in \cite{Fogli:2009ce}).

\subsection{Origin of flavour hierarchies?}

The specific patterns of hierarchies in fermion masses and
mixings observed in the lepton (and quark) flavour sector
suggest a theoretical explanation in terms of additional
(flavour) symmetries. A well-known example is the
Froggat-Nielsen approach \cite{Froggatt:1978nt}, which postulates an additional
(spontaneously broken) $U(1)$ symmetry together
 with heavy fermionic messenger fields and a particular
choice of $U(1)$ charges for the different families.
Other popular approaches are discrete (non-abelian) flavour
symmetries (see e.g.\ \cite{Ma:2009wt}), which lead to particular textures in the fermion Yukawa
matrices, or models with extra spatial dimensions (ED), where the different
fermion families are diplaced along the extra dimension and the mass hierarchies
are explained by the imperfect overlap of the corresponding wave function profiles. 
For an overview of different models, see \cite{Raidal} and
references therein.

\section{Lepton-flavour violation}

As already mentioned, LFV decays provide particularly sensitive
probes of NP at the TeV scale, and the predicted correlations 
between various observables may help to distinguish different  models,
like SUSY, ED, Little Higgs \ldots \
LFV observables may be classified as follows: (i)
dipole transitions, which induce the decays 
$\mu \to e\gamma$, $\tau \to \mu(e)\gamma$;
(ii) 4-lepton transitions, $\mu \to 3e$, $\tau \to 3\mu$ etc.,
which also receive contributions from dipole operators via virtual photons;
(iii) transitions involving 2 leptons and 2 quarks, which
induce LFV hadronic decays, and $\ell$-$\ell'$ conversion in nuclei.\footnote{
Of similar importance for the phenomenological analysis of NP models 
are flavour-diagonal transitions, like the anomalous magnetic
moment of the muon, $(g-2)_\mu$, or lepton electric dipole moments (EDMs).}

\begin{figure}[t]
\begin{center} \footnotesize
 \begin{tabular}{ccc} 
 \quad LLLL & \quad LLRR & \quad dipole (LR) \\
\includegraphics[width=0.3\textwidth]{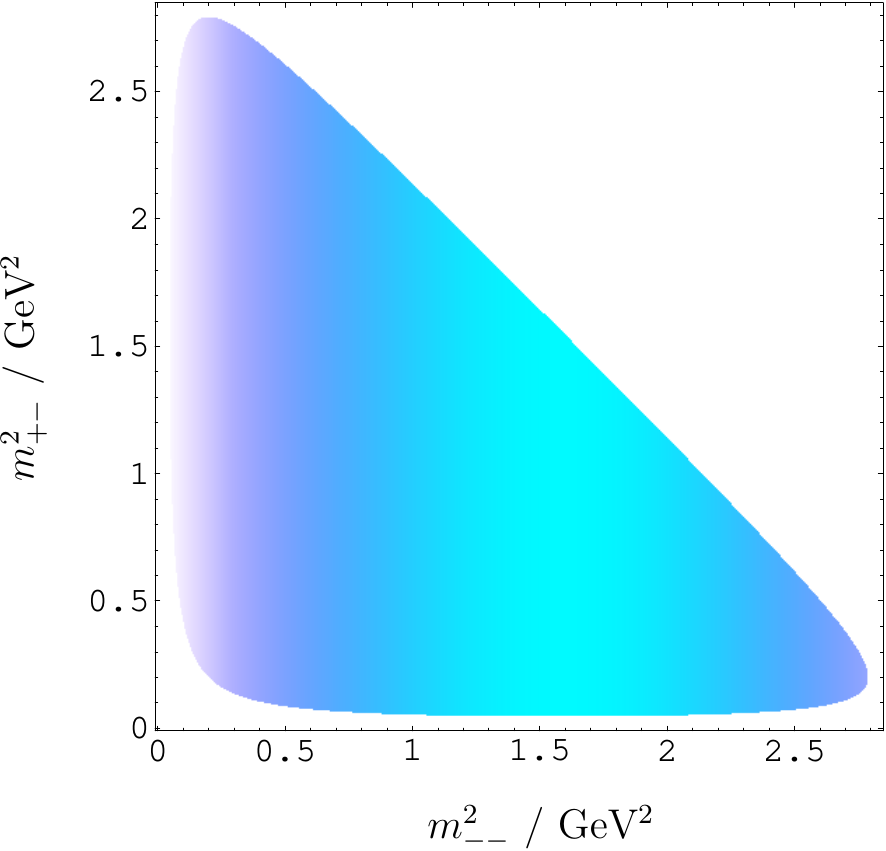}  &
\includegraphics[width=0.3\textwidth]{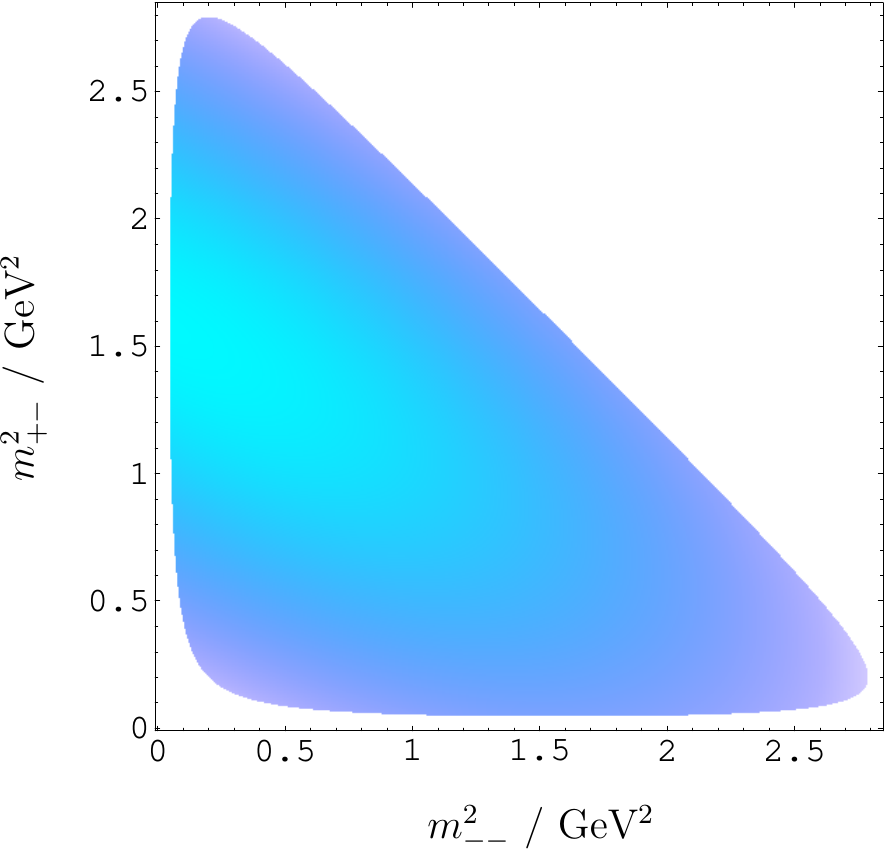}  &
\includegraphics[width=0.3\textwidth]{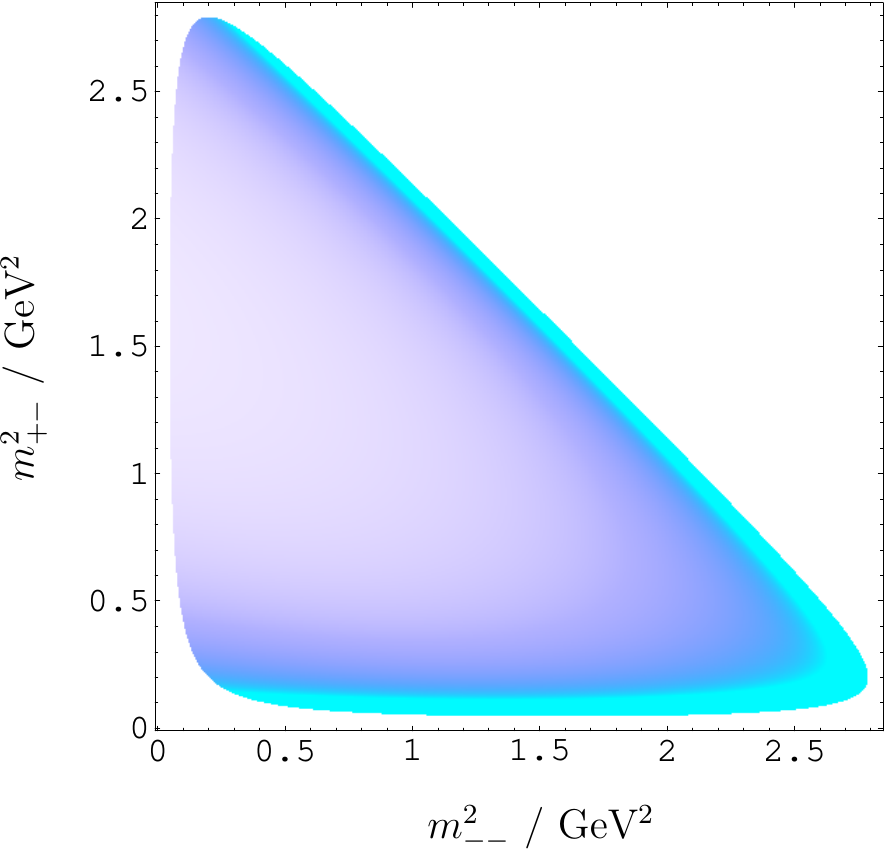} 
\end{tabular}
\end{center}
\caption{\label{fig:ps} 
Examples for phase space distributions in $\tau \to 3\mu$ induced by
4-lepton operators of different chirality (LLLL, LLRR), or by virtual photons from
dipole operators (LR) \cite{Dassinger:2007ru}. 
In the Dalitz plots the vertical (horizontal) axis 
refers to the invariant mass of a muon pair of opposite (same) charge. 
}
\end{figure}

While dipole transitions provide direct constraints on
the coefficients of the corresponding operators $\ell \sigma^{\mu\nu} \ell' F_{\mu\nu}$,
the analysis of 3-body leptonic decays is complicated by the fact that
different chiralities in 4-lepton operators and the virtual photon contributions
from dipole operators lead to different phase-space distributions
\cite{Dassinger:2007ru}.
Depending on the dominance of one or the other operator, different regions
in the Dalitz plot show different sensitivity on NP parameters
(examples are shown in Fig.~\ref{fig:ps}). This has
to be taken into account for a model-independent analysis of experimental data.
Finally, LFV transitions of class (iii) require hadronic matrix elements
as non-perturbative input.
In the following, we will present some illustrative examples based on
specific NP predictions, and also comment on model-independent
approaches based on minimal-flavour violation (MFV).

\subsection{LFV phenomenology in specific models}

\begin{figure}[t]
\begin{center}  
\includegraphics[width=0.41\textwidth]{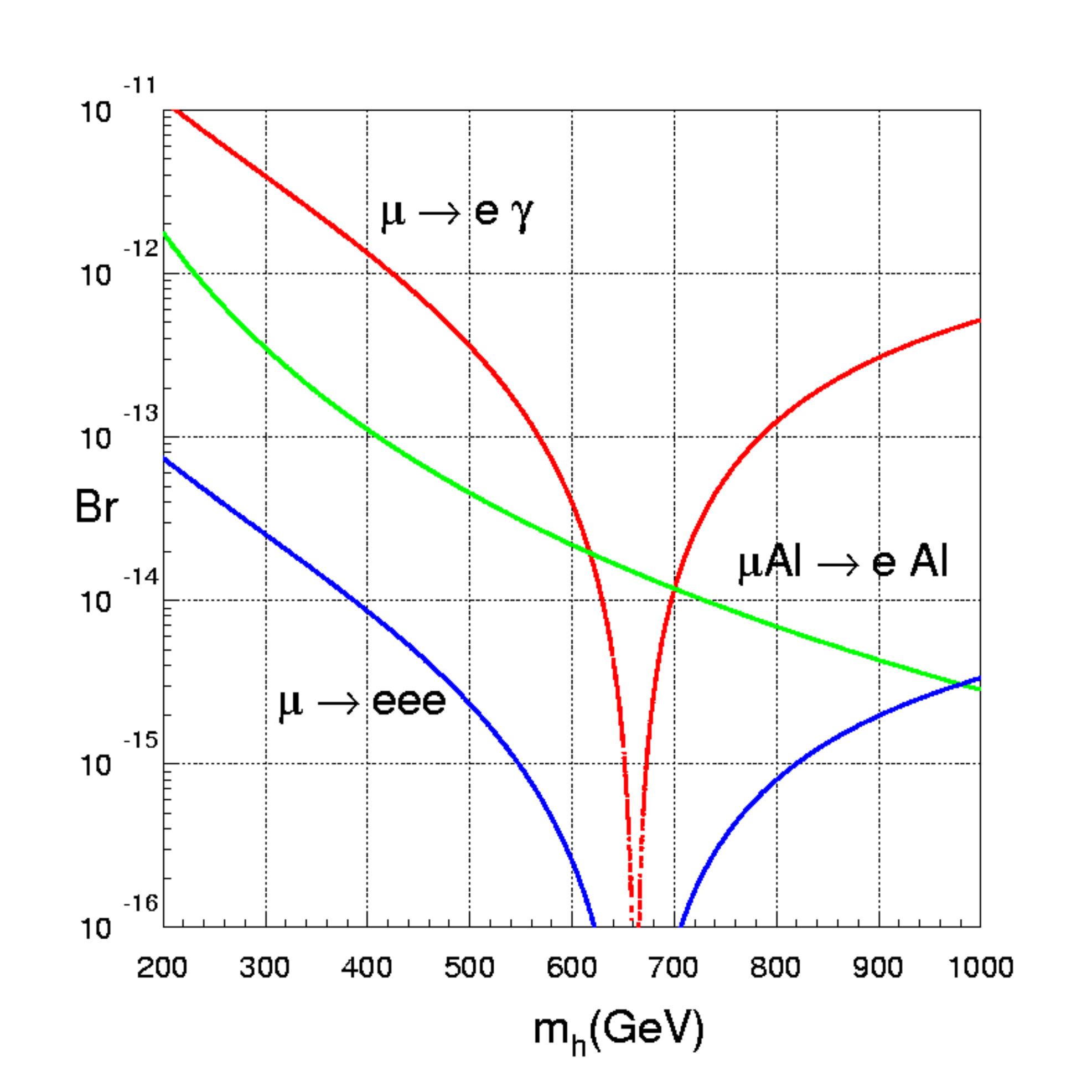} \qquad 
\includegraphics[width=0.41\textwidth]{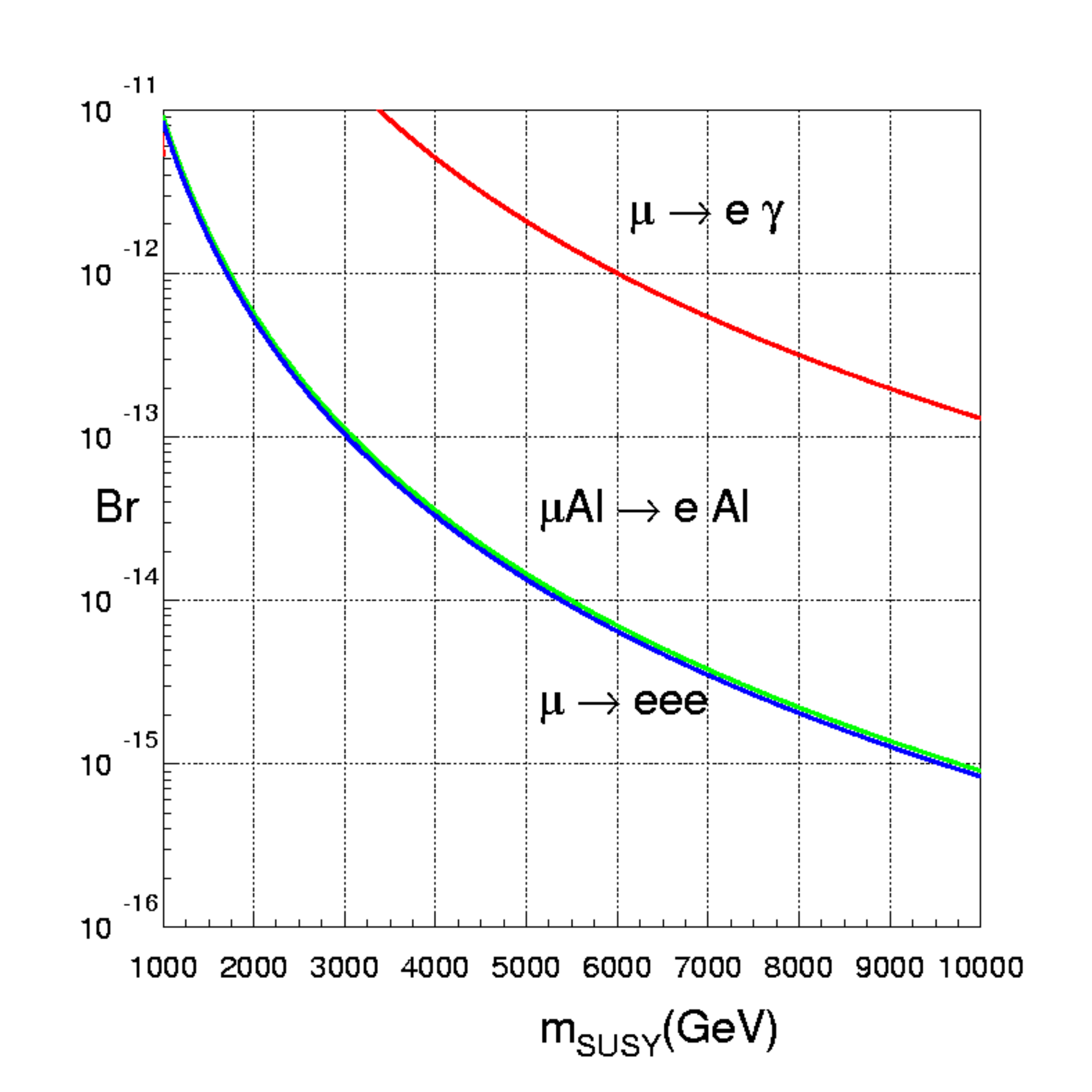}
\end{center}
\caption{\label{fig:Higgsm}
Branching ratios of $\mu\to e\gamma$, $\mu\to eee$ and $\mu {\rm Al}\to e {\rm Al}$.
Left: For Higgs-mediated LFV as a function of the Higgs boson mass $m_h$.
Right: For gaugino-mediated LFV as a function of a common SUSY mass $m_{\rm SUSY}$. In both cases $\tan\beta=50$ and $\delta^{21}_{LL}=10^{-2}$. Figures taken from \cite{Paradisi:2006jp}.}
\end{figure}

SUSY extensions of the SM
generically provide many new sources for LFV and CP-violation,
and the phenomenological consequences have been extensively 
studied in the literature. 
Depending on see-saw parameters and on the
(small) misalignment between fermions 
and sfermions in the soft SUSY-breaking sector, one can study correlations
between different LFV decays.  
In a generic SUSY see-saw framework, one can derive inequalities like
\cite{Ibarra:2008uv}
\begin{align}
 {\rm BR}[\mu \to e\gamma] \gtrsim C \times 
 {\rm BR}[\tau \to \mu\gamma] \, {\rm BR}[\tau \to e\gamma] \,,
\end{align}
where $C$ is a constant that can be calculated for a given
set of SUSY parameters.
More explicit correlations, e.g.\ 
between $\mu\to e\gamma$ and $\tau \to \mu\gamma$
\cite{Antusch:2006vw}, can only be obtained 
by specifying additional assumptions about (otherwise unobservable) 
Majorana neutrino parameters.
Furthermore, depending on the SUSY parameters, LFV decays can be either
dominated by gaugino-mediated or by Higgs-mediated flavour
transitions, where the latter arise from loop-induced non-holomorphic
Higgs couplings.\footnote{Notice that the possible strength of
Higgs-induced flavour transitions in SUSY  is also constrained 
from quark decays measured e.g.\ in $B \to \tau \nu_\tau$ or $B \to X_s\gamma$.}
In the two cases, the observables considered in Fig.~\ref{fig:Higgsm}
show quite different correlations.
Finally, interesting correlations between $\mu \to e\gamma$ and 
flavour-diagonal observables  ($(g-2)_\mu$
and leptonic EDMs) have been explored in \cite{Hisano:2009ae}.

\begin{figure}[t]
 \begin{center}
\includegraphics[width=0.57\textwidth]{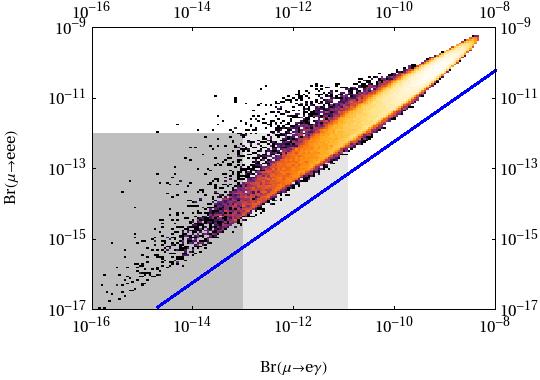}
\end{center}
\caption{\label{fig:lht} 
$ {\cal B}[\mu \to 3 e]$ vs.\ $ {\cal B}[\mu \to e\gamma]$
in LHT  for $f=1~$TeV and 
$ 300~{\rm GeV} \leq M_{H_i}^\ell \leq 1.5~{\rm TeV}$
from \cite{Blanke:2009am}. The lower line denotes the
contribution from the dipole operator via $\mu \to e\gamma^*$,
and the light (dark) grey areas the present (foreseen) exp.\
bounds.}
\end{figure}

Another class of SM extensions with interesting LFV effects
are  ``Littlest Higgs Models'' with $T$-parity (LHT).
Besides new gauge bosons (which can be detected at the LHC), these
models contain new doublets of mirror leptons (and quarks) with masses
of order TeV, which may induce LFV rates that exceeds the SM case by orders
of magnitude. Phenomenological predictions depend on the LHT scale parameter, $f$,
the masses of the mirror leptons, $M_{H_i}^\ell$, the 3 mixing angles among the
mirror leptons $\theta_{ij}^\ell$ and 3 new (Dirac) phases $\delta_{ij}^\ell$.
In Fig.~\ref{fig:lht} we show as an example the correlation between
$ {\cal B}[\mu \to 3 e]$ and $ {\cal B}[\mu \to e\gamma]$ for $f=1~$TeV, and 
$ 300~{\rm GeV} \leq M_{H_i}^\ell \leq 1.5~{\rm TeV}$ as calculated in 
\cite{Blanke:2009am}. Two important conclusions can be drawn:
First, the mirror leptons must be quasi-degenerate
   and/or the mixings have to be very hierarchical in order 
   to fulfill the present bounds on the individual decays.
Second, the considered LHT scenarios, where 3-lepton decays (and also $\mu$-$e$ conversion)
are dominated by $Z^0$-penguin and box diagrams,
can be clearly distinguished from the MSSM, where dipole operators
(or Higgs-boson induced effects) play the dominant role.
For a comparison, see 
Table~\ref{tab:comp} \cite{Blanke:2009am} (for a discussion of LFV effects
in minimal see-saw models, see \cite{Kamenik}).
\begin{table}[t]
\caption{\label{tab:comp} Comparison of LFV transitions in LHT and MSSM:
Branching fractions for 3-lepton decays and $\mu$-$e$ conversion, relative
to the corresponding radiative decays 
(from \cite{Blanke:2009am}).}
\begin{center}  
\begin{minipage}{0.95\textwidth}
\begin{tabular*}{\textwidth}{@{\extracolsep{\fill}}l|ccc}
\hline \hline 
ratio & LHT  & MSSM (dipole) & MSSM (Higgs) \\\hline
$\frac{Br(\mu^-\to e^-e^+e^-)}{Br(\mu\to e\gamma)}$  & \hspace{.8cm} 0.02\dots1\hspace{.8cm}  & $\sim6\cdot10^{-3}$ &$\sim6\cdot10^{-3}$  \\
$\frac{Br(\tau^-\to e^-e^+e^-)}{Br(\tau\to e\gamma)}$   & 0.04\dots0.4     &$\sim1\cdot10^{-2}$ & ${\sim1\cdot10^{-2}}$\\
$\frac{Br(\tau^-\to \mu^-\mu^+\mu^-)}{Br(\tau\to \mu\gamma)}$  &0.04\dots0.4     &$\sim2\cdot10^{-3}$ & $0.06\dots0.1$ \\\hline
$\frac{Br(\tau^-\to e^-\mu^+\mu^-)}{Br(\tau\to e\gamma)}$  & 0.04\dots0.3     &$\sim2\cdot10^{-3}$ & $0.02\dots0.04$ \\
$\frac{Br(\tau^-\to \mu^-e^+e^-)}{Br(\tau\to \mu\gamma)}$  & 0.04\dots0.3    &$\sim1\cdot10^{-2}$ & ${\sim1\cdot10^{-2}}$\\
$\frac{Br(\tau^-\to e^-e^+e^-)}{Br(\tau^-\to e^-\mu^+\mu^-)}$     & 0.8\dots2.0   &$\sim5$ & 0.3\dots0.5\\
$\frac{Br(\tau^-\to \mu^-\mu^+\mu^-)}{Br(\tau^-\to \mu^-e^+e^-)}$   & 0.7\dots1.6    &$\sim0.2$ & 5\dots10 \\\hline
$\frac{R(\mu\text{Ti}\to e\text{Ti})}{Br(\mu\to e\gamma)}$  & $10^{-3}\dots 10^2$     & $\sim 5\cdot 10^{-3}$ & $0.08\dots0.15$ \\
\hline\hline 
\end{tabular*}
\end{minipage}
\end{center}
\end{table}

%%%%%%%%%%%%%%%

\subsection{Minimal flavour violation}

As already mentioned, the non-observation of LFV decays puts
severe constraints on NP parameters. For instance, allowing for
generic coupling constants in front of effective operators,
the bound on $\mu \to e\gamma$ would already set a lower bound on
the NP scale,  $\Lambda_{\rm LFV} > 10^5$~TeV. 
In order to avoid ad-hoc fine-tuning of parameters, one could 
introduce the concept of minimal flavour violation in the
lepton sector (MLFV \cite{Cirigliano,Smith}).
Using an effective-theory approach, one relates the NP flavour
coefficients to the SM ones by considering the flavour matrices as
vacuum expectation values (VEVs) of spurion fields. 
Compared to the quark sector, additional
complications arise, because the neutrino masses
themselves already imply a (model-dependent) extension of the SM.
Therefore the specification of the neutrino field content and
the mechanism for the generation of neutrino masses should be
considered part of the effective-theory construction, on top of the
MFV hypothesis. In particular, the scale for 
lepton-\emph{number} violation $\Lambda_{L}$
should be distinguished from the LFV scale, $\Lambda_{\rm LFV}$.

Focusing on the minimal extension of the SM, 
the flavour group in the lepton sector is given by
$U(3)_L \times U(3)_{E_R}$, i.e.\ independent unitary
transformations of left-handed doublets and right-handed singlets.
The flavour symmetry is broken by the charged-lepton Yukawa matrices
and the flavour matrix in front of the effective dim-5 operator (\ref{dim5}).
\begin{align}
\mbox{Yukawa:} \quad Y_E \sim (3,\bar 3)_{1,-1} \,,
\qquad
\mbox{Dim-5:} \quad g_\nu \sim (\bar 6,1)_{2,0} \,.
\end{align}
In the mass eigenbasis for charged leptons, the matrix $g_\nu$ can
be expressed in terms of the neutrino masses and the PMNS mixing matrix,
\begin{align}
 g_\nu = \frac{\Lambda_L}{v^2} \, U_{\rm PMNS}^* \, {\rm diag}[m_{\nu}] \, U_{\rm PMNS}^\dagger \,.
\end{align}
In MLFV, effective 2- and 4-lepton operators for flavour transitions
are now constructed in such a way that
they are formally invariant under the flavour group (and the SM gauge group),
which is achieved by inserting appropriate powers of $g_\nu$ and $Y_E$ with
overall (flavour-independent) coupling constants of order 1.
A complete set of operators can be found in \cite{Cirigliano}.
\begin{table}[t]
\caption{\label{tab:break} Partial breaking of the lepton-flavour symmetry
group by the atmospheric mass squared difference for normal or inverted
hierarchy \cite{Feldmann:2008av}.}
\begin{center} 
\begin{tabular}{lcl}
 \hline\hline 
hierarchy & symmetry & approx.\ spurion VEV
\\
\hline 
 normal & $ \begin{array}{l}  SU(3)_L \times U(1)_L \\
  \to  U(2)_L \times Z_2 \end{array} $ &
$ 
 \langle g_\nu \rangle \simeq \left( \begin{array}{ccc}
       0 & 0 & 0 \\
       0 & 0 & 0 \\
       0 & 0 & 1          
                \end{array}
\right) \cdot   \frac{\Lambda_{\slashed L} \sqrt{\Delta m^2_{\rm atm}}}{v^2} $
\\
\hline 
 inverted &
 $\begin{array}{l} SU(3)_L \times U(1)_L \\ \to SO(2)_L \times U(1)_{L_3}
  \end{array}$ &
$
 \langle g_\nu \rangle \simeq \left( \begin{array}{ccc}
       1 & 0 & 0 \\
       0 & 1 & 0 \\
       0 & 0 & 0          
                \end{array}
\right) \cdot  \frac{\Lambda_{\slashed L}  \sqrt{\Delta m^2_{\rm atm}}}{v^2}$ 
\\
\hline\hline
\end{tabular}
\end{center}
\end{table}
The leading effect due to large mass-squared difference 
$\Delta m_{\rm atm}^2$ observed in atmospheric neutrino oscillations
 can be singled out by using a non-linear representation of MLFV
\cite{Feldmann:2008av}. %[TF/Mannel, arXiv:0806.0717]}
Depending on the neutrino mass hierarchy, this also implies
a particular (partial) breaking of the lepton flavour symmetry,\footnote{This
can be viewed as the first step of a sequence of flavour symmetry breaking,
as discussed for the quark sector in \cite{Feldmann:2008ja}.}
see Table~\ref{tab:break}. The solar mass differences, as well as
the mixing parameters in $U_{\rm PMNS}$ are then associated to
spurion fields of the residual flavour symmetry.

\begin{figure}[t] 
\begin{center}
 \includegraphics[width=0.54\textwidth]{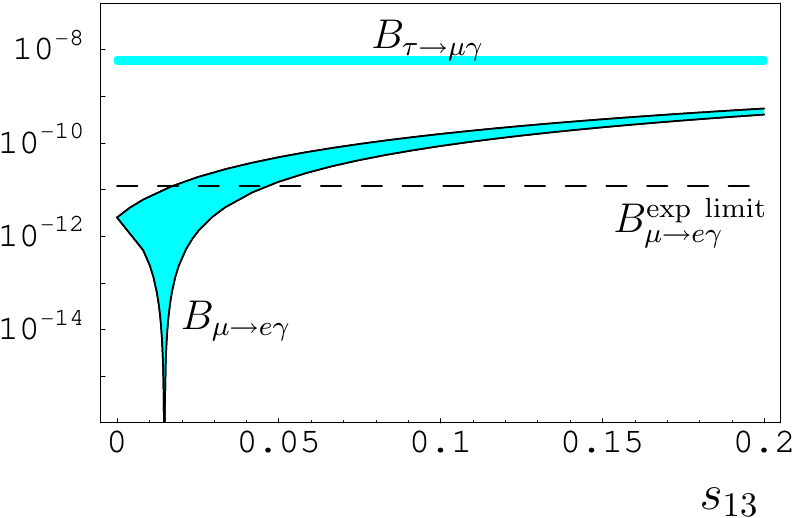}
\end{center}
\caption{\label{fig:mlfv}
MLFV predictions for ${\cal B}(\mu \to e\gamma)$ and ${\cal B}(\tau \to \mu\gamma)$
as a function of $\sin\theta_{13}$, using $\Lambda_{\slashed L} = 10^{10} 
\cdot \Lambda_{\rm LFV}$. Figure from \cite{Cirigliano}.}
\end{figure}

A comprehensive phenomenological study of the MLFV scenario (with minimal field
content) in \cite{Cirigliano} revealed that 
LFV decay rates are sizeable only if the scales for lepton-number violation
and flavour violation are clearly separated.
% $\Lambda_{\slashed L} \gg \Lambda_{\rm LFV}$.
For instance, a sizeable rate ${\cal B}(\mu \to e\gamma) > 10^{-13}$ requires  
$\Lambda_{\slashed L} > 10^9 \cdot \Lambda_{\rm LFV}$.
On the other hand,
$\Lambda_{\slashed L}$ drops out in \emph{ratios} of LFV observables,
for instance MLFV predicts
${\cal B}(\mu \to e\gamma)/{\cal B}(\tau \to \mu\gamma) \sim  10^{-2}-10^{-3}$,
see also Fig.~\ref{fig:mlfv}. Among others, this implies better experimental
prospects to observe $\mu\to e\gamma$ than $\tau\to\mu\gamma$ in the near
future. Also the LFV decays of light hadrons are typically very small
in MFV scenarios. On should stress that the above conclusions are
relaxed if one extends the field content. At the same time, however, the 
MLFV becomes less predictive because of the increased number
of flavour parameters. For more details, see \cite{Cirigliano}.

\section{Summary}

The observation of neutrino oscillations requires to
extend the Standard Model. 
While the see-saw mechanism for neutrino masses and
leptogenesis scenarios point towards   
lepton-\emph{number} violating new physics near/below
the GUT scale, many generic models also allow for
lepton-\emph{flavour} violating effects at/above
the TeV scale.
The exploration of the lepton-flavour sector can thus
be viewed as complementary to precision measurements of the 
CKM parameters and rare quark decays and to the direct search for
new particles and interactions at hadron colliders in
the LHC era.

\subsection*{Acknowledgements}

I would like to thank M.~Krawczyk, H.~Czyz and M.~Misiak for 
organizing a very stimulating and interesting workshop.
I am also grateful to A.~Ibarra and P.~Paradisi for a critical reading of
the manuscript and helpful comments.

\end{document}